\documentclass[12pt]{aipproc}
\usepackage{epsfig,graphics}

\newcommand{\tr}{\rm tr \,}

\newcommand{\TableHeader}[0]{\begin{tabular}{|c|c|c|c|c|c|} \hline (I,S)  &  Kanal &  Lage &  |g|  & Lage  &|g| \\}
\def\rescale{\fontsize{6}{1}}

\layoutstyle{6x9}
\begin{document}
\title{Coupled-channel study of \\baryon resonances with charm}
\keywords{coupled channel systems, hadronic molecules, chiral
dynamics}

\author{J. Hofmann}{address={Gesellschaft f\"ur Schwerionenforschung (GSI)\\
Planck Str. 1, 64291 Darmstadt, Germany}}
\author{M.F.M. Lutz}{address={Gesellschaft f\"ur Schwerionenforschung (GSI)\\
Planck Str. 1, 64291 Darmstadt, Germany}}
\begin{abstract}
Identifying a zero-range exchange of vector
mesons as the driving force for the s-wave scattering of pseudo-scalar mesons
off the  baryon ground states, a rich spectrum of molecules
is formed. We argue that chiral symmetry and
large-$N_c$ considerations determine that part of the interaction which generates the spectrum.
A bound state with exotic quantum numbers is predicted at mass 2.78 GeV. It
couples strongly to the $(\bar D_s\,N),(\bar D \,\Lambda ),(\bar D\,\Sigma)$ channels.
A further charm minus-one system is predicted at mass 2.84 GeV as a result of
$(\bar D_s \Lambda), (\bar D\,\Xi)$ interactions.
The two  so far observed s-wave baryons with charm one are
recovered. We argue that the $\Lambda_c(2880)$ is not a s-wave state.
In addition to those states we predict the existence of about ten narrow s-wave
baryon states with masses below 3 GeV. A triplet of crypto-exotic states decaying
dominantly into channels with an $\eta'$ is obtained with masses 4.24 GeV and 4.44 GeV.
\end{abstract}

\classification{12.38 Cy,12.38 Lg, 12.39 Fe}

\maketitle

\section{Introduction}

It is the purpose of the present talk to review recent progress on
the nature of charmed baryon systems \cite{Hofmann:Lutz:05}.
Empirically the QCD spectrum of charmed baryon states is poorly
studied so far. It is important to correlate the properties of
charmed states to those firmly established, applying a unified and
quantitative framework. Our strategy is to extend previous works
\cite{LK04-charm,LK05} that performed a coupled-channel study of
the s-wave scattering processes where a Goldstone boson hits an
open-charm baryon ground state. The spectrum of
$J^P=\frac{1}{2}^-$ and $J^P=\frac{3}{2}^-$ molecules obtained in
\cite{LK04-charm,LK05} is quite compatible with the so far very
few observed states. We mention that analogous computations
successfully describe the spectrum of open-charm mesons with
$J^P=0^+$ and $1^+$ quantum numbers \cite{KL04,HL04}. These
developments were driven by the conjecture that meson and baryon
resonances that do not belong to the large-$N_c$ ground state of
QCD should be viewed as hadronic molecular states
\cite{LK02,LWF02,LK04-axial,Granada,Copenhagen}. Extending those
computations to include $D$- and $\eta_c$-mesons in the
intermediate states leads to additional baryon states
\cite{Hofmann:Lutz:05,Mo:80,Rho:Riska:Scoccola:92,Min:Oh:Rho:95,Oh:Kim:04}.
The results of \cite{LK04-charm,LK05} were based on the leading
order chiral Lagrangian, that predicts unambiguously the s-wave
interaction strength of Goldstone bosons with open-charm baryon
states in terms of the pion decay constant. Including the light
vector mesons as explicit degrees of freedom in a chiral
Lagrangian gives an interpretation of the leading order
interaction in terms of the zero-range t-channel exchange of light
vector mesons \cite{Weinberg:68,Wyld,Dalitz,sw88,Bando:85}. The
latter couple universally to any matter field in this type of
approach. Given the assumption that the interaction strength of
$D$- and $\eta_c$-mesons with the baryon ground states is also
dominated by the t-channel exchange of the light vector mesons, we
are in a position to perform a quantitative coupled-channel study
of charmed baryon resonances.

The interaction strengths of the channels
that drive the resonance generation are predicted by chiral and large-$N_c$ properties of QCD.
The spectrum of $J^P=\frac{1}{2}^-$ molecules obtained is amazingly rich of structure.
Interesting results are obtained in the charm minus one sector. Exotic states
with strangeness minus one and two are predicted at mass 2.78 GeV and 2.84 GeV.
In the charm one sector we recover the $\Lambda_c(2593)$ as a narrow state coupling
strongly to the $(D\,N)$ and $(D_s\,\Lambda)$ states. The $\Xi_c(2790)$ is interpreted as a
bound state of the $(\bar K \,\Sigma_c),(\eta \,\Xi'_c)$ system. We argue that the
$\Lambda_c(2880)$ discovered by the CLEO collaboration can not be a s-wave state.
About ten additional narrow s-wave states are predicted in this sector with masses below 3 GeV.

\section{Coupled-channel interactions}

We study the interaction of pseudoscalar mesons with the
ground-state baryons composed out of u,d,s,c quarks. The
pseudoscalar mesons that are considered in this work can be
grouped into multiplet fields $\Phi_{[9]}, \Phi_{[\bar 3]}$ and
$\Phi_{[1]}$, corresponding to the Goldstone-bosons together with
the $\eta'$ meson, the $D$ mesons and the $\eta_c$ meson.

The baryon states are collected into SU(3) multiplet fields $B_{[8]}, B_{[6]}, B_{[3]}$ and
$B_{[\bar 3]}$ with charm 0,\,1,\,1 and 2.
We construct the interaction of the mesons and baryon fields with the vector mesons
\begin{eqnarray} \label{LM}
{\mathcal L}^{\rm{SU(4)}}_{\rm{int,M}} &=& {\textstyle{i\over 4}}\,g\,{\tr}\Big(
 \big[  (\partial_\mu\,\Phi_{[16]})\,,\Phi_{[16]} \big]_- V^\mu_{[16]}\Big) \nonumber\\
 &+&{\textstyle{1\over 4}}\,g \sum_{i,j,k,l=1}^4\,
\bar B^{[20]}_{ijk} \,\gamma^\mu\,\Big( V_{\mu,\,l}^{[16],k}\,B^{ijl}_{[20]}
+ 2\, V_{\mu,\,l}^{[16],j}\,B^{ilk}_{[20]}\Big)\,,
\end{eqnarray}
written down in terms of SU(4)-symmetric fields.
Within the hidden local symmetry model \cite{Bando:85} chiral symmetry is recovered.
It is acknowledged that chiral symmetry does not constrain the coupling
constants involving the SU(3) singlet part of the fields. The latter can, however, be
constrained  by a large-$N_c$ operator analysis \cite{DJM}.
At leading order in the $1/N_c$ expansion the OZI rule \cite{OZI} is predicted.
We emphasize that the combination of chiral and large-$N_c$ constraints
determine all couplings
of the light vector mesons consistent with the SU(4)-symmetric interaction (\ref{LM}).

The prediction of the vertex (\ref{LM}) can be tested against the
decay pattern of the D meson. From the empirical branching ratio
\cite{PDG04} we deduce $g = 10.4 \pm 1.4$, which is compared with
the estimate $g \simeq 5.8 $ that follows from the $\rho$ decay.
We observe a small SU(4) breaking pattern. Based on this result
one may expect (\ref{LM}) to provide magnitudes for the coupling
constants reliable within a factor two. The precise values of the
coupling constants will not affect the major results of this work.
This holds as long as those coupling constants range in the region
suggested by (\ref{LM}) within a factor three.

We consider the s-wave scattering of the pseudo-scalar mesons fields
off the baryon fields. The scattering kernel is approximated
by the t-channel vector meson exchange force defined by
(\ref{LM}), where we apply the formalism developed in \cite{LK02,LK04-axial}.
The scattering kernel has the form
\begin{eqnarray}
K^{(I,S,C)}(\bar q,q;w) = -\frac{1}{4}\,\sum_{V \in [16]}\,
\frac{C^{(I,S,C)}_V}{t-m^2_V} \,\Big( \frac{\ \makebox[0mm]{$\bar q$} \makebox[0mm]{{ \big /}}\ \! + \ \makebox[0mm]{$q$} \makebox[0mm]{{ \big /}}\ \!}{2}
- (\bar q^2-q^2)\,\frac{ \ \makebox[0mm]{$\bar q$} \makebox[0mm]{{ \big /}}\ \! - \ \makebox[0mm]{$q$} \makebox[0mm]{{ \big /}}\ \! }{2\,m_V^2}\Big)\,,
\label{def-K}
\end{eqnarray}
with the initial and final meson 4-momenta $q_\mu$ and $\bar q_\mu$ and $t=(\bar q-q)^2$.
In (\ref{def-K}) the scattering is projected onto sectors with conserved isospin (I),
strangeness (S) and charm (C) quantum numbers.

The first term of the interaction kernel matches corresponding expressions predicted
by the leading order chiral Lagrangian if we put $t=0$ in (\ref{def-K}) and use the common
value for the vector-meson masses suggested by the KSFR relation.
The second term in (\ref{def-K}) is formally of chiral order $Q^3$ for
channels involving Goldstone bosons. Numerically it is a minor correction but nevertheless it is
kept in the computation.

\section{S-wave resonances with charm}

The properties of a collection of resonance states with $C = \pm
1$ that are generated dynamically are listed in Tab.
\ref{tab:result}. The masses of the $(I,S)=(\frac{1}{2},-1)$ and
$(0,-2)$ states with $C=-1$ are predicted at 2.78 GeV and 2.84 GeV
respectively. For details of the choices of parameters we refer to
\cite{Hofmann:Lutz:05}. We point out that none of the coupling
constants fixed by a SU(4) assumption affect the spectrum. As a
consequence of the OZI rule only the t-channel exchange of the
light-vector mesons contribute.

\begin{table}[t]
\rescale \setlength{\tabcolsep}{1.8mm}
\setlength{\arraycolsep}{3.2mm}
\renewcommand{\arraystretch}{1.35}
\begin{tabular}{|ll|c|c|l|c|c|}
\hline &$ (\,I,\phantom{+}S)$  &  $M_R [\rm MeV]$ &
$ \rm Width [MeV]$  &$ (\,I,\phantom{+}S)$ & $M_R [\rm MeV]$ & $\rm Width [MeV]$  \\
\hline
$C=-1:$&$(\frac12,-1)$      & $2780$ & $0$
&$(0,-2)$   & $2838$ & $0$\\
\hline $C=+1:$&$(\frac12,\phantom{+}1)$  & $2932$ & $6.9$
&$(\frac12,\phantom{+}1)$   & $2892$ & $0.6$\\
\cline{3-7}
&$(0,\phantom{+}0)$     & $2593$ & $0.05$
&$(0,\phantom{+}0)$     & $2815$ & $0.0001$\\
\cline{3-7}
&$(0,\phantom{+}0)$     & $3023$ & $19$
&$(0,\phantom{+}0)$     & $3068$ & $22$ \\
\cline{3-7}
&$(0,\phantom{+}0)$     & $4238$ & $7.3$
&$(1,\phantom{+}0)$     & $2620$ & $1.4$ \\
\cline{3-7}
&$(1,\phantom{+}0)$     & $2992$ & $18$
&$(\frac12,-1)$         & $2672$ & $0.06$ \\
\cline{3-7}
&$(\frac12,-1)$         & $2793$ & $16$
&$(\frac12,-1)$         & $2759$ & $0.9$ \\
\cline{3-7}
&$(\frac12,-1)$         & $3104$ & $43$
&$(\frac12,-1)$         & $4443$ & $9$ \\
\cline{3-7}
&$(\frac32,-1)$         & $3052$ & $15$
&$(0,-2)$           & $2805$ & $0$ \\
\cline{3-7}
&$(0,-2)$           & $2927$ & $0$
&$(0,-2)$           & $2953$ & $0$ \\
\cline{3-7}
&$(1,-2)$           & $3815$ & $5$
&\ \ --- & & \\
\hline
\end{tabular}
\caption{\mbox{Spectrum of $J^P=\frac{1}{2}^-$ baryons with charm
$C=\pm 1$.}} \label{tab:result}
\end{table}

Charm systems with $C=+1$ are quite intriguing since the channels
which have either a charmed baryon or a charmed meson are
comparatively close in mass. Unfortunately, at present there is
very little empirical information available on open-charm s-wave
resonances. Only two states, the $\Lambda_c(2593)$ and $
\Xi_c(2790)$ \cite{PDG04} are discovered so far. We claim that the
$\Lambda_c(2880)$ observed by the CLEO collaboration \cite{Artuso}
can not be a s-wave resonance. This will be substantiated below.

In previous coupled-channel computations the effect of
the light pseudo-scalar mesons as they scatter off charmed baryons was studied
\cite{LK04-charm,LK05}. We confirm the striking prediction of such
computations which suggest the existence of strongly bound $\bar 3, 6$ but also weakly
bound $\bar 3, 6,\overline{15}$ systems. These multiplets are formed by
scattering the octet of Goldstone bosons off the  baryon anti-triplet and sextet.
\begin{eqnarray}
&& 8 \otimes \bar 3= \bar 3 \oplus 6 \oplus \overline{15} \,,\qquad
8 \otimes 6= \bar 3 \oplus 6 \oplus \overline{15}\oplus 24\,.
\label{8times3}
\end{eqnarray}
For $8 \otimes \bar 3$ scattering chiral dynamics predicts attraction in anti-triplet, sextet
and repulsion in the $\overline {15}$-plet \cite{LK04-charm}. For $8 \otimes 6$ scattering
attraction is foreseen in anti-triplet, sextet, $\overline{15}$-plet with decreasing strength.
Further multiplets are generated by the scattering of the anti-triplet mesons of the octet
baryons. The decomposition is given already in (\ref{8times3}). In this case we find attraction in
the anti-triplet, sextet and the $\overline{15}$-plet. If we switch off the t-channel forces defined by the exchange
of heavy vector mesons the three types of resonances discussed above do not communicate with each
other. This is a direct consequence of the chiral SU(3) symmetry. It forbids the transition of an anti-triplet baryon into a
sextet baryon under the radiation of a light vector meson. Since the exchange of heavy vector
mesons, which mixes the three kinds of states, is largely suppressed, the SU(4) assumption in (\ref{LM}) has
a very minor effect on the resonance spectrum.

The anti-triplet states are identified most easily in the $(0,0)$ sector.
The narrow state at 2.593 GeV couples strongly to the anti-triplet mesons. It has properties
amazingly consistent with the $\Lambda_c(2593)$ \cite{PDG04}. The empirical width is
$3.6^{+2.0}_{-1.3}$ MeV. This narrow state is almost degenerate in mass with a chiral excitation
of the triplet baryons \cite{LK04-charm,LK05}.
It decays dominantly into the $\pi\, \Sigma_c$ channel giving it a width of about 50 MeV.
A further narrow state at 2.815 GeV is the second chiral excitation of the
anti-triplet baryons \cite{LK04-charm,LK05} in this sector. Since it couples strongly
to the $\eta\,\Lambda_c(2285)$ channel, one should not associate this state with the
$\Lambda_c(2880)$ detected by the CLEO collaboration \cite{Artuso} via
its decay into the $\pi\, \Sigma_c(2453)$ channel.
The narrow total width of the observed state of smaller than 8 MeV \cite{Artuso} appears
inconsistent with a large coupling of that state to the open $\eta\,\Lambda_c$ channel.
The chiral excitation of the 6 baryon is quite broad in
this sector with mass around 2.65 GeV coupling strongly to the $\pi \Sigma_c$ channel.
Most spectacular is the $(\frac{1}{2},-1)$ sector in which we predict 4 narrow states below 4
GeV. The particle data group reports a state $\Xi_c(2790)$ with a decay width
smaller than 15 MeV. It is naturally identified with the chiral excitation of the sextet baryon of
mass 2.79 GeV and width 16 MeV. This state couples strongly to
the $\bar K\,\Sigma_c$ and $\eta \,\Xi'_c$ channels.

Crypto-exotic states with $cc\bar c$ content are
formed by the scattering of the $3$-plet mesons with $C=-1$ off the triplet baryons with $C=2$:
\begin{eqnarray}
3 \otimes 3 = \bar 3 \oplus  6 \,,
\end{eqnarray}
where we predict strong attraction in the anti-triplet sector
only. The associated narrow states have masses ranging from 4.1
GeV to 4.4 GeV. Like in the case of the crypto-exotic states in
the zero-charm sectors these states decay preferably into channels
involving the $\eta'$ \cite{Hofmann:Lutz:05}. Depending on the
parameter set the widths is large about 200-250 MeV or down to a
few MeV.

\newpage

\section{Summary}

We reported on a coupled-channel study of s-wave baryon resonances
with charm $\pm 1$. A rich spectrum is predicted in terms of a
t-channel force defined by the exchange of light vector mesons.
All relevant coupling constants are obtained from chiral and
large-$N_c$ properties of QCD. Less relevant vertices related to
the t-channel forces induced by the exchange of charmed vector
mesons  were estimated by applying SU(4) symmetry.

We confirm the expectation of
\cite{Gignoux:Silvestre-Brac:Richard:87,Lipkin:87} that
penta-quark type states exists with charm minus one. Binding is
predicted only in systems with strangeness minus one and minus
two. In the charm one sector several so far unobserved narrow
states are predicted.

Our predictions can be tested experimentally in part by existing collaborations
like SELEX or BELLE. We urge the QCD lattice community to perform unquenched simulations
in order to verify or disprove the existence of the predicted states. Such studies are of great
importance since they will shed more light on how confinement is realized in nature. The
central question - what are the most relevant degrees of freedom responsible for the formation of
resonances in QCD - can be studied best in systems involving light and heavy quarks
simultaneously. The spectrum of open-charm baryons could be a topic of interest for the FAIR
project at GSI.

\end{document}